\documentclass{article}
\usepackage{epsfig}
\usepackage{amsmath}
\usepackage{color}
\newtheorem{theorem}{Theorem}[section]
\newtheorem{lemma}{Lemma}[section]

\newcommand{\blackslug}{\penalty 1000\hbox{
    \vrule height 8pt width .4pt\hskip -.4pt
    \vbox{\hrule width 8pt height .4pt\vskip -.4pt
          \vskip 8pt
      \vskip -.4pt\hrule width 8pt height .4pt}
    \hskip -3.9pt
    \vrule height 8pt width .4pt}}

\newenvironment{proof}{\vspace{1mm} \noindent {\sc Proof.}$\;$\rm}{\qed}
\newcommand{\qed}{\hspace*{\fill}\blackslug}
\def\boxit#1{\vbox{\hrule\hbox{\vrule\kern4pt
 \vbox{\kern1pt#1\kern1pt}
\kern2pt\vrule}\hrule}}
\begin{document}

\title{\bf Lower Bounds for the Domination Numbers of Connected Graphs without Short Cycles}

\author{Yinglei Song \\
School of Electronics and Information Science\\
Jiangsu University of Science and Technology\\
Zhenjiang, Jiangsu 212003, China\\
syinglei2013@163.com\\
}
\date{}
\maketitle

\begin{abstract}

\noindent In this paper, we obtain lower bounds for the domination numbers of connected graphs with girth at least $7$. We show that the domination number of a connected graph with girth at least $7$ is either $1$ or at least $\frac{1}{2}(3+\sqrt{8(m-n)+9})$, where $n$ is the number of vertices in the graph and $m$ is the number of edges in the graph. For graphs with minimum degree $2$ and girth at least $7$, the lower bound can be improved to $\max{\{\sqrt{n}, \sqrt{\frac{2m}{3}}\}}$, where $n$ and $m$ are the numbers of vertices and edges in the graph respectively. In cases where the graph is of minimum degree $2$ and its girth $g$ is at least $12$, the lower bound can be further improved to $\max{\{\sqrt{n}, \sqrt{\frac{\lfloor \frac{g}{3} \rfloor-1}{3}m}\}}$.
\end{abstract}


\section{Introduction}

Let $G=(V,E)$ be a graph, $D \subseteq V$ is a {\it dominating set} in $G$ if each vertex in $V-D$ is adjacent to at least one vertex in $D$. A dominating set $D$ is a {\it minimum dominating set} if $G$ does not contain a dominating set of cardinality less than that of $D$. In graph theory, the {\it domination number} of a graph is the cardinality of a minimum dominating set in the graph.

Extensive research has been conducted to obtain upper bounds for  the domination numbers of graphs. In \cite{ore}, it is shown that the domination number of a connected graph is at most the half of the number of vertices in the graph. For graphs of minimum degree 2, it is shown in \cite{blank} and \cite{mccuaig} that the domination numbers of almost all such graphs are at most $\frac{2n}{5}$, where $n$ is the number of vertices in a graph. In \cite{mccuaig,randerath,xu}, it is shown that these upper bounds are sharp. In \cite{reed}, an upper bound of $\frac{3n}{8}$ is established for any graph of minimum degree $3$, where $n$ is the number of vertices in the graph.

The {\it girth} of a graph is the length of the shortest cycle in the graph. In \cite{kostochka}, it is shown that the domination number of any connected cubic graph of order at least 8 and girth $g$ is at most $(\frac{1}{3}+\frac{8}{3g^{2}})n$, where $n$ is the number of vertices in the graph. In \cite{kawarabayashi}, the domination numbers of 2-edge connected cubic graphs with girth $3k$ are studied and it is shown that an upper bound of $(\frac{1}{3}+\frac{1}{9k+3})n$ holds for the domination numbers of such graphs, where $n$ is the number of vertices in such a graph.

In \cite{brigham}, an upper bound of $\lceil \frac{n}{2}-\frac{g}{6} \rceil$ is obtained for graphs with minimum degree $2$ and girth $g$. Similar results on the upper bounds of such graphs are also shown in \cite{volkmann1,volkmann2,rautenbach}. In \cite{lowenstein}, graphs with minimum degree 2 and girth at least $5$ are studied and an upper bound of
$(\frac{1}{3}+\frac{2}{3g})n$ is established for the domination number of such graphs, where $n$ is the number of vertices in the graph and $g$ is the girth of the graph. Results on both lower and upper bounds for the independent domination numbers of graphs of girth $6$ are provided in \cite{haviland}.

In this paper, we establish lower bounds for the domination numbers of connected graphs with girth at least $7$. Using a technique based on graph partition, we show that the domination number of a connected graph with girth at least $7$ is either $1$ or at least $\frac{1}{2}(3+\sqrt{8(m-n)+9})$, where $m$ and $n$ are the numbers of edges and vertices in the graph respectively. For graphs with minimum degree $2$ and girth at least $7$, we show that the lower bound can be improved to $\max{\{\sqrt{n}, \sqrt{\frac{2m}{3}}\}}$. When the girth $g$ of the graph is at least $12$, we show that the lower bound can be further improved to $\max{\{\sqrt{n}, \sqrt{\frac{\lfloor \frac{g}{3} \rfloor-1}{3}m}\}}$.

\section{Results}
\subsection{A Partition Theorem}
The lower bounds for domination numbers are obtained by partitioning a graph into disjoint vertex subsets. Given a graph $G=(V,E)$, we use $\gamma (G)$ to denote the domination number of $G$. A vertex subset $D \subseteq V$ is {\it outer-dominated} if there exists a vertex $u$ such that $u$ is not in $D$ and $u$ is adjacent to each vertex in $D$. We show that vertices in $G$ can be partitioned into $\gamma (G)$ disjoint vertex subsets such that each vertex subset contains a vertex from a minimum dominating set in $G$ and is not outer-dominated.

\begin{theorem}
\label{th1}
\rm
Let $G=(V,E)$ be a graph and $\gamma(G)$ is the domination number of $G$. Vertices in $G$ can be partitioned into $\gamma(G)$ disjoint vertex subsets $S_1, S_2, \cdots, S_{\gamma(G)}$ such that each $S_i$ ($1 \leq i \leq \gamma(G)$) contains a vertex from a minimum dominating set in $G$ and is not outer-dominated.

\begin{proof}
Since the domination number of $G$ is $\gamma(G)$, $G$ contains a minimum dominating set $D$ that contains $\gamma(G)$ vertices. We use $u_1, u_2, \cdots, u_{\gamma(G)}$ to denote the vertices in $D$. We then partition the vertices in $G$ with the following algorithm.
\begin{enumerate}
\item{Initialize $S_1, S_2, \cdots, S_{\gamma(G)}$ to be
$\{u_1\}, \{u_2\}, \cdots, \{u_{\gamma(G)}\}$ respectively;}
\item{if there exists a vertex $u$ that is not included in any of $S_1, S_2, \cdots, S_{\gamma(G)}$, we find all vertices in $D$ that are adjacent to $u$ in $D$ and arbitrarily pick a vertex $u_i$ from them;}
\item{update $S_i$ to be $S_i \cup \{u\}$;}
\item{go back to step 2 if there exists a vertex that is not included in any of $S_1, S_2, \cdots, S_{\gamma(G)}$, otherwise continue to execute step 5;}
\item{color each vertex in $G$ to be red;}
\item{if there exists a vertex $v$ and two different subsets $S_i$, $S_j$ such that $v \in S_i$ and $v$ is adjacent to each vertex in $S_j$, move $v$ from $S_i$ to $S_j$ and color $v$ to be green; otherwise continue to execute step 8;}
\item{go back to step 6;}
\item{output $S_1, S_2, \cdots, S_{\gamma(G)}$ as the result of partition.}
\end{enumerate}

We show that the above algorithm terminates and outputs a disjoint partition of vertices in $G$. We first show that none of vertices in $D$ are moved and colored to be green by the algorithm. To see this, we consider $u_i \in D$. From step 2, $u_i$ is adjacent to all red vertices in $S_i$. From step 6, any green vertex later added to $S_i$ is adjacent to $u_i$. If $u_i$ is also adjacent to all vertices in a different subset $S_j$, $D-\{u_j\}$ is a dominating set in $G$ and it contains $\gamma(G)-1$ vertices only. This contradicts the fact that $D$ is a minimum dominating set. Such a subset $S_j$ thus does not exist. From step 6, we conclude that $u_i$ is never moved during the execution of the algorithm.

We then show that a vertex is moved from one subset to another for at most once. In other words, a green vertex is never moved from a subset to another one. To see this, consider a green vertex $w \in S_i$, we claim that $w$ is adjacent to all other vertices in $S_i$ since from step 6, $w$ is adjacent to all red vertices in $S_i$ and each green vertex added later to $S_i$ is also adjacent to $w$. If $w$ is also adjacent to all vertices in another subset $S_j$, $(D-\{u_i, u_j\})\cup \{w\}$ is a dominating set in $G$ and contains only $\gamma(G)-1$ vertices. This is contradictory to the fact that $D$ is a minimum dominating set. Such a vertex subset $S_j$ thus does not exist and $w$ is never moved again after it is moved to $S_i$.

Since $G$ contains $|V|$ vertices in total and a vertex is moved for at most once, step 6 is executed for at most $|V|$ times. The algorithm thus halts and outputs a disjoint partition of vertices in $G$. It is straightforward to see that each subset in the partition is not outer-dominated. The theorem thus follows.
\end{proof}
\end{theorem}

\subsection{The Lower Bounds}

We now consider the domination numbers of connected graphs that are of girth at least $7$. Based on Theorem \ref{th1}, we show that for any such graph that contains $n$ vertices and $m$ edges, the domination number is either $1$ or at least $\frac{1}{2}(3+\sqrt{8(m-n)+9})$. For graphs that are of minimum degree $2$ and girth at least $7$, an improved lower bound $\max{\{\sqrt{n}, \sqrt{\frac{2m}{3}}\}}$ can be obtained.

\begin{theorem}
\label{th2}
\rm
Let $G=(V,E)$ be a connected graph of girth at least $7$. The domination number of $G$ is either $1$ or at least $\frac{1}{2}(3+\sqrt{8(m-n)+9})$, where $m$ and $n$ are the numbers of edges and vertices in $G$ respectively. The domination number of $G$ is at least $\max{\{\sqrt{n}, \sqrt{\frac{2m}{3}}\}}$ if the minimum degree of $G$ is $2$.

\begin{proof}
The domination number of $G$ is $1$ when it is a star. We thus assume $G$ is not a star. We use $\gamma(G)$ to denote the domination number of $G$. $G$ contains a minimum dominating set $D$ of cardinality $\gamma(G)$. We use $u_1, u_2, \cdots, u_{\gamma(G)}$ to denote the vertices in $D$. From Theorem \ref{th1}, vertices in $G$ can be partitioned into $\gamma(G)$ disjoint subsets $S_1, S_2, \cdots, S_{\gamma(G)}$ such that $u_i \in S_i$ for each $1 \leq i \leq \gamma(G)$ and $S_i$ is not outer-dominated.

We consider the subgraph induced on $S_i$ in $G$. Note that $u_i \in S_i$ and all other vertices in $S_i$ are adjacent to $u_i$. The subgraph induced on $S_i$ is a star centered at $u_i$ since otherwise a cycle of length $3$ exists in $G$, which is contradictory to the fact the girth of $G$ is $7$. In addition, $S_i$ contains at least $2$ vertices since $S_i$ is not outer-dominated.

We now divide the edges in $G$ into two disjoint subsets $I_1$ and $I_2$. Namely, $I_1$ is the set of edges contained in subgraphs induced on $S_1, S_2, \cdots, S_{\gamma(G)}$. An edge in $I_1$ is an {\it inner-edge} in $G$. $I_2$ is the set of edges that are not in $I_1$. An edge in $I_2$ is an {\it intra-edge} in $G$.

Since the subgraph induced on $S_i$ is a star centered at $u_i$, $I_1$ contains at most $n-\gamma(G)$ edges. For two arbitrary subsets $S_i$ and $S_j$, the number of intra-edges that join a vertex from $S_i$ and another one from $S_j$ is at most $1$, since otherwise a cycle of length at most $6$ exists in the subgraph induced on $S_i \cup S_j$, which contradicts the fact that the girth of $G$ is at least $7$. The total number of intra-edges is thus at most $\frac{1}{2}\gamma(G)(\gamma(G)-1)$. Since $m=|I_1|+|I_2|$, the following inequality holds for $\gamma(G)$.
\begin{eqnarray}
m &=& |I_1|+|I_2| \\
  &\leq &  n-\gamma(G)+\frac{1}{2}\gamma(G)(\gamma(G)-1) \\
  & \leq & \frac{1}{2}\gamma^{2}(G)-\frac{3}{2}\gamma(G)+n
\end{eqnarray}
Since $G$ is not a star, $\gamma(G) \geq 2$. From the above inequality, we obtain
\begin{equation}
\gamma(G) \geq \frac{1}{2}(3+\sqrt{8(m-n)+9})
\end{equation}

We then consider the case where the minimum degree of $G$ is $2$. Since the subgraph induced on $S_i$ is a star centered at $u_i$ and the degree of each vertex in $S_i-\{u_i\}$ is at least $2$, each vertex in $S_i-\{u_i\}$ is connected to at least another vertex not in $S_i$ by an intra-edge. However, since the number of intra-edges between any pair of subsets is at most $1$, the number of intra-edges incident to a vertex in $S_i$ is at most $\gamma(G)-1$. The number of vertices in $S_i$ is thus at most $\gamma(G)$. Since $n=\sum_{i=1}^{\gamma(G)}|S_i|$, we immediately obtain
\begin{eqnarray}
n & = & \sum_{i=1}^{\gamma(G)}|S_i| \\
  & \leq & \gamma^{2}(G)
\end{eqnarray}
where the inequality is due to the fact that $|S_i| \leq \gamma(G)$. From the above inequality, we obtain $\gamma(G) \geq \sqrt{n}$. On the other hand, since $|S_i| \leq \gamma(G)$, the number of inner-edges is at most $\gamma(G)(\gamma(G)-1)$. From $m=|I_1|+|I_2|$, we obtain
\begin{eqnarray}
  m & = & |I_1|+|I_2| \\
    & \leq & \gamma(G)(\gamma(G)-1)+\frac{1}{2}\gamma(G)(\gamma(G)-1) \\
    & \leq & \frac{3}{2}\gamma^{2}(G) \\
\end{eqnarray}
The above inequality leads to $\gamma(G) \geq \sqrt{\frac{2}{3}m}$. The theorem thus follows.
\end{proof}
\end{theorem}

We now show that an improved upper bound can be obtained when the graph is of minimum degree $2$ and its girth is at least $12$. We need the following lemma to bound the number of edges in a graph of girth $g$ from above.

\begin{lemma}
\label{lm1}
\rm
Let $G=(V,E)$ be a graph of girth at least $g$ ($g \geq 3$), $G$ contains at most $\frac{1}{g-1}n^{2}$ edges, where $n$ is the number of vertices in $G$.

\begin{proof}
We consider all vertex subsets that contain $g-1$ vertices in $G$. Since the girth of $G$ is at least $g$, the subgraph induced on each of such subsets is a forest, the number of edges contained in such a subgraph is thus at most $g-2$. The sum of the edges in all such subgraphs is at most $(g-2){n \choose g-1}$.

On the other hand, each edge in $G$ is counted for ${n-2 \choose g-3}$ times in the sum of the edges in all such subgraphs. We there immediately obtain
\begin{equation}
   |E|{n-2 \choose g-3} \leq (g-2){n \choose g-1}
\end{equation}
From the above inequality, we can immediately obtain
\begin{eqnarray}
   |E| & \leq & \frac{(g-2){n \choose g-1}}{{n-2 \choose g-3}} \\
       &  = & \frac{n(n-1)}{g-1} \\
       &  \leq & \frac{1}{g-1}n^{2}
\end{eqnarray}
The lemma thus follows.
\end{proof}
\end{lemma}

\begin{theorem}
\label{th3}
\rm
Let $G=(V,E)$ be a graph of girth $g$ ($g \geq 12$) and minimum degree $2$, the domination number of $G$ is at least $\max{\{\sqrt{n}, \sqrt{\frac{\lfloor \frac{g}{3} \rfloor-1}{3}m}\}}$, where $m$ and $n$ are the numbers of edges and vertices
in $G$.

\begin{proof}
We use $\gamma(G)$ to denote the domination number of $G$. As in the proof of Theorem \ref{th2}, $G$ can be partitioned into $\gamma(G)$ disjoint vertex subsets $S_1, S_2, \cdots, S_{\gamma(G)}$ based on a minimum dominating set $D=\{u_1, u_2, \cdots, u_{\gamma(G)}\}$ in $G$. The subgraph induced on each $S_i$ ($1 \leq i \leq \gamma(G)$) is a star centered at $u_i$. Similarly, $I_1$ and $I_2$ are the sets of inner-edges and intra-edges respectively. Since $g \geq 12$, $|I_1| \leq \gamma(G)(\gamma(G)-1)$.

We use a vertex $r_i$ to represent each $S_i$ and a graph $H$ of $\gamma(G)$ vertices can be constructed to describe the relationships among the $\gamma(G)$ vertex subsets. Two vertices $r_i$ and $r_j$ are joined by an edge in $H$
if there exists an intra-edge that connects a vertex in $S_i$ to another one in $S_j$ in $G$. We claim that $H$ is a graph of girth at least $\lfloor \frac{g}{3} \rfloor$ since otherwise $G$ contains a cycle of length at most $3(\lfloor \frac{g}{3} \rfloor-1) <g$, which contradicts the fact that the girth of $G$ is $g$.We use $E(H)$ to denote the number of edges in $H$ and $d_i$ to denote the degree of $r_i$ in $H$. Since each vertex in $G$ is of degree at least $2$, $S_i$ contains at most $d_i$ inner-edges. The total number 
of inner-edges is thus at most $\sum_{i=1}^{\gamma(G)}d_{i}=2E(H)$. On the other hand, the number of intra-edges is at most $E(H)$
 Let $l= \lfloor \frac{g}{3} \rfloor$. From Lemma \ref{lm1}, $E(H)$ is at most $\frac{1}{l-1}\gamma(G)(\gamma(G)-1)$. Since $m=|I_1|+|I_2|$, we obtain
\begin{eqnarray}
 m &=& |I_1|+|I_2| \\
   & \leq &  3E(H) \\
   & \leq & \frac{3}{l-1}\gamma^{2}(G)
\end{eqnarray}
From the above inequality, we obtain
\begin{eqnarray}
    \gamma(G) & \geq & \sqrt{\frac{l-1}{l}m} \\
              & = & \sqrt{\frac{\lfloor \frac{g}{3} \rfloor-1}{3}m}
\end{eqnarray}
The theorem thus follows.
\end{proof}
\end{theorem}

We need to point out here that the lower bound in Theorem \ref{th3} can be further improved when $g=12, 13$ or $14$ due to a well known fact that a triangle free graph contains at most $\frac{n^2}{4}$ edges, where $n$ is the number of vertices in the graph.

\section{Conclusions}

In this paper, we study the lower bounds for the domination numbers of connected graphs with girth at least $7$. Based on a partition based technique, we obtain a lower bound for the domination numbers of such graphs. In addition, we show that the lower bound can be improved if the minimum degree of the graph is at least $2$. In cases where the girth of the graph is at least $12$, we show that the lower bound can be further improved.

A closer analysis can show that the lower bound we have obtained is sharp when the girth of the graph is $7$. However, since the upper bound on the edge number in Lemma \ref{lm1} is not sharp, it is very likely that the lower bound in Theorem \ref{th3} is not sharp either. Therefore, whether improved lower bounds can be obtained for the domination numbers of graphs with large girths or not is an interesting problem and constitutes an important part of our future work.

\end{document}